\documentclass[epj]{svjour}
\usepackage{amsfonts}
\usepackage{amssymb}
\usepackage{graphicx}
\newcommand{\ds}{\displaystyle}

\newcommand{\qvec}{\mbox{\boldmath $q$}}
\newcommand{\kvec}{\mbox{\boldmath $k$}}

\newcommand{\tauvec}{\mbox{\boldmath $\tau$}}
\newcommand{\pivec}{\mbox{\boldmath $\pi$}}

\newcommand{\nrmsq}{\mbox{$\langle r^2\rangle_n$}}
\newcommand{\etal}{\mbox{\it et al.}}

\begin{document}
\title{The electric form factor of the neutron and its chiral content}
\author{Murat M. Kaskulov  \and Peter Grabmayr
}
\institute{Physikalisches Institut, Universit\"at  T\"ubingen,
		 D-72076 T\"ubingen, Germany}
\date{\today}
\abstract{
 Considering the nucleon as a system of confined valence quarks surrounded by
  pions we derive a Galster-like parameterization of the neutron electric form
  factor $G_E^n$.  Furthermore, we show that the proposed parameterization can
  be linked to properties of the pion cloud. By this, the high quality data
  for the pion form factor can be used in predictions of $G_E^n$ in the low
  $Q^2$ region, where the direct double polarization measurements are not
  available.
} 
\maketitle

The electromagnetic (e.m.) form factors of the nucleons contain all the
 information on the internal nucleon structure and in particular they are very
 sensitive to the details of the interaction between the valence quarks.
 Already the first analyses by Hofstadter~\etal~\cite{Hofstadter} demonstrated
 that the proton electric form factor can be described by a dipole type form
\begin{equation}
\label{eq:dipole}
G_D(Q^2) = (1+Q^2/\Lambda^2)^{-2}
\end{equation}
Using the canonical $\Lambda^2_D$=0.71~GeV$^2$, $G_E^p$ is reproduced up to
 four momentum transfer $Q^2\sim$~1~GeV$^2$.  Also the magnetic nucleon form
 factors, $G_M^p$ and $G_M^n$, are reproduced reasonably well.  Recently, new
 precise polarization experiments and new
 analyses~\cite{Brash,Kelly,FriedrichW,Gao,KG} have brought clear evidence for
 the deviations of the form factors from the simple dipole form at high $Q^2$.
 Slight deviations from the dipole form at low $Q^2$ have been attributed to
 the underlying pionic and quarkonic structure of the
 nucleons~\cite{FriedrichW}.

The neutron electric form factor $G_E^n$ is the most uncertain one due to its
 vanishing net charge and the absence of free neutron targets. The internal
 structure of the neutron is also reflected by a finite charge
 radius~\cite{Fermi,Foldy}.  Understanding the measured
 \nrmsq=-0.115$\pm$0.003~\cite{Kopecky} is still an interesting goal, since
 the contributing Foldy term~\cite{Foldy} of $3F_2(0)/2M_N^2$=-0.126~fm$^2$
 almost equals the measured \nrmsq\ value. At the same time, a recent analysis
 by Isgur~\cite{IsgurNeutron} indicates that the Foldy term does not really
 explain the neutron charge radius and its charge distribution, because in
 leading order of the relativistic approximation to the constituent quark
 model (CQM) the Foldy term is canceled exactly by a contribution to the Dirac
 form factor~$F_1$.  This result was confirmed by Ref.~\cite{Leinw} where it
 was argued that both, the success of CQM in reproducing the ratio of the
 proton to neutron magnetic moments and the success of the Foldy term in
 reproducing the observed charge radius of the neutron, are coincidental.

The extraction of the nonzero charge form factor $G_E^n$ from elastic electron
 scattering off the deuteron is rather model
 dependent~\cite{Hofstadter,Galster,Platchkov}. In recent years $G_E^n$ is
 being studied successfully by double polarization experiments~\cite{Becker}
 and the most recent parameterizations in the range up to $Q^2\sim$~1~GeV$^2$
 seem to converge~\cite{Kelly,FriedrichW,Gao,Pgbuchmann,Sick}.  The so called
 Galster~\cite{Galster} parameterization of $G_E^n$ is based on the same
 dipole form $G_D(Q^2)$ as for the proton. In oder to account for the
 condition $G_E^n(Q^2=0)$=0 required by the vanishing charge of the neutron,
 $G_D$ is multiplied by an appropriate function. Out of four trial
 functions~\cite{Galster},
\begin{equation}
\label{eq:Galster}
	G_E^n(Q^2) = \frac{a_G\tau}{1+b_G\tau }~G_D(Q^2)
\end{equation}
served best, where $\tau=Q^2/4M_n^2$ and $M_n$=0.939~GeV$^2$ is the neutron
 mass. Originally~\cite{Galster}, the parameter~$a_G$ was set equal to
 $-\mu_n$ and the fit to the data resulted in $b_G$=5.6. More recent fits,
 e.g. Ref.~\cite{FriedrichW}, obtain $a\sim$1.73 which reproduces the measured
 root mean square radius of the neutron and determines $b$=4.62.

Theoretically, up to now the Galster parameterization has no particular
 theoretical justification and is considered as a purely empirical
 description, i.e. $a_G$ and $b_G$, are fitting parameters.  In this paper we
 wish to show that the parameterization of the neutron electric form factor
 $G_E^n$ can be derived directly under certain assumptions about the pionic
 content of the nucleon.  We found that all parameters of
 Eq.~(\ref{eq:Galster}) can be fixed by other experiments and are connected
 with the existence of pions in the nucleons. In order to disentangle shape
 and magnitude of $G_E^n$ we propose the use of the parameter $a'$ which is
 defined as $a_G=a'\cdot{b_G}$.  The parameter $b_G$ can be related to the
 pion electromagnetic form factor~$F_\pi$ and $a'$ to the spectroscopic
 strength of the pions (number of pions) in the nucleon.  With this
 prescription, reasonable parameters entering in Eq.~(\ref{eq:Galster}) are
 obtained supporting the proposed interpretation. Using ``our'' Galster-like
 parameterization we are also able to relate the data for the pion e.m. form
 factor $F_{\pi}$ as provided by electroproduction experiments to the neutron
 electric form factor $G_E^n$. By this procedure, a high quality
 representation of $G_E^n$ can be obtained in the low momentum transfer
 region, where the direct double polarization data are still not available and
 would have large corrections due to final state interaction.

By now it is well established that the pion cloud plays an important role in
 understanding the variety of electromagnetic and hadronic properties of
 nucleons in the low-energy, non-perturbative region of
 QCD~\cite{Kaskulov:2003bg,ThomasWeise}. The coupling of the pion field to the
 nucleon quark core and the resulting pion-loop (pion cloud) corrections are
 important ingredients of the so-called chiral quark models
 ($\chi$QM)~\cite{ThomasWeise} where the pions are the Goldstone bosons
 generated by spontaneous breaking of $SU(2)_R{\times}SU(2)_L$ chiral
 symmetry. {Easily the scheme can be generalized to the octet of light
 pseudo-scalar mesons ($\pi$, K, $\eta$) as provided by
 $SU(3)_R{\times}SU(3)_L$}~\cite{Lyub}. Formally, the pion cloud can be introduced into
 the nucleon structure by the perturbative expansion of the nucleon Fock
 space. This procedure is reflected in the two-component structure of the
 nucleon wave function $\Psi_N = (\Psi_{3q},\Psi_{3q + \pi})$, where the first
 component, $\Psi_{3q}$, represents the bare nucleon consisting of valence
 quarks, and the second one, $\Psi_{3q + \pi}$, is the quark core dressed by a
 pion cloud which is mainly responsible for the soft physics. The presence of
 a soft pion cloud as an actual dynamical degree of freedom is crucial in
 understanding the neutron electric form factor $G_E^n$. In $\chi$QM's at the
 one-pion loop level, the neutron charge form factor is a first order effect
 of the pion cloud and originates mainly from the Fock component of the
 neutron wave function consisting of a $\pi^-$ cloud and a positively charged
 core of confined quarks. This physical picture gives a natural explanation of
 the nonvanishing charge distribution inside the neutron which otherwise, like
 in the simplest version of non-relativistic quark model (NRQM) with three
 valence quarks only, results in zero~\cite{Kaskulov:2003bg}.

\begin{figure}[t]
\begin{center}
\includegraphics[clip=true,width=0.75\linewidth]{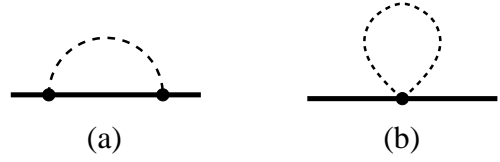}
\caption{\label{fig:Pion_Loop}
One-pion loop self-energy diagrams.
 The intermediate state in (a) can be a nucleon or a $\Delta$.
 The tadpole (Hartree) diagram in (b) vanishes due to its isospin structure.
}
\end{center}
\end{figure}

Consider the effective $SU(2)_R{\times}SU(2)_L$ Lagrangian of $\chi$QM which
 can be written formally as~\cite{ThomasWeise}
\begin{equation}
\label{L}
\mathcal{L} = \mathcal{L}_{\pi}^{(2)} + \mathcal{L}_{\pi qq}^{(1)}
\end{equation}
where the mesonic Lagrangian~$\mathcal{L}_{\pi}^{(2)}$ of lowest-order in the
 derivative expansion is given by the nonlinear $\sigma$ model and the
 $\mathcal{L}_{\pi q q}^{(1)}$ is an effective pion-quark Lagrangian
\begin{eqnarray}
\label{Lpiqq}
\mathcal{L}_{\pi q q}^{(1)} = 
-\frac{1}{4f_\pi^2} \bar{\psi} \gamma^\mu{\tauvec}\psi
[ \pivec \times \partial_{\mu}  \pivec ]
- \frac{f_{\pi qq}}{m_\pi} 
\bar{\psi}\gamma^\mu\gamma_5{\tauvec} \psi\partial_\mu\pivec ~~~
\end{eqnarray}
where $\psi$ and $\pivec$ are a quark and pion fields, respectively, and the
 coupling constant~$f_{\pi{qq}}$ characterizes the strength of the pion-quark
 interaction.  The coupling of the pion field to the core of confined quarks,
 Eq.~(\ref{Lpiqq}), results in pion-loop corrections as provided by the
 self-energy pion-loop diagrams shown in Fig.~\ref{fig:Pion_Loop} already on
 the hadronic level. Note that the underlying quark substructure of diagrams
 is reflected in appearance of hadronic form factors as shown by filled
 circles.  The tadpole diagram (Fig.~\ref{fig:Pion_Loop}b) vanishes due to its
 isospin structure and Fig.~\ref{fig:Pion_Loop}a) is characterized by the loop
 integral $-i\Sigma_{\alpha}(E)$ (see its explicite expression in
 Ref.~\cite{ThomasWeise}), where the intermediate states $\alpha$ are a
 $N(939)$ or a $\Delta(1232)$.  Fig.~\ref{fig:Pion_Loop}a introduces the
 pionic degrees of freedom into the nucleon structure and mathematically
 requires the renormalization of the total nucleon wave function
$ \ds| \Psi_N \rangle^{R} = Z_{2}^{-1/2} | \Psi_N \rangle$,
where the constant $Z_2$ is given by:
\begin{equation}
Z_2\equiv 1 - \sum_{\alpha} {\partial\Sigma_{\alpha} (E)}/{\partial E} |_{E=M_N}
\end{equation}
 and describes the probability of finding the ``physical'' nucleon in its bare
 three-quark valence Fock state. It also guarantees
 $\ds^R\langle\Psi_N|\Psi_N\rangle^R=1$. The value $Z_2$ can be related to the
 pionic content of the nucleon, namely to the pion spectroscopic factor,
 which represents the number of pions in the nucleon $\ds S_{\pi} =
 \sum_\alpha S^\alpha_{\pi} = (Z_2 - 1)/Z_2$ or
\begin{equation}
\label{eq:genpi}
S^\alpha_\pi = - \frac{1}{Z_{2}}
\frac{\partial\Sigma_\alpha(E)}{\partial E} \Big|_{E=M_N} 
\end{equation}

\begin{figure}[t]
\begin{center}
\includegraphics[clip=true,width=0.75\linewidth]{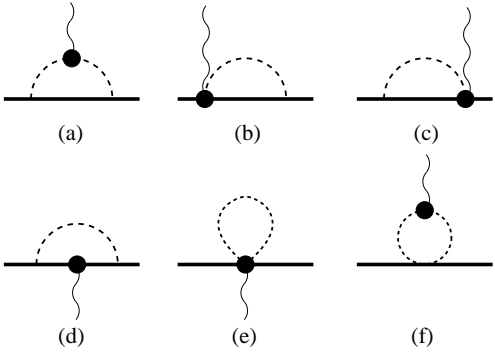}
\caption{\label{fig:Currents}
E.m. currents at one-pion loop level.}
\end{center}
\end{figure}

The explicit introduction of the additional degrees of freedom in the nucleon
 structure will change its properties as compared to expectations based on
 simple quark models in which the nucleon is described as a system of three
 valence quarks only. For example, the coupling of the
 e.m. field~$\mathcal{A}^{\mu}$ to the conserved e.m. currents due to a quark
 charge~$\mathcal{Q}$=$[1/3+\tau_3]/2$ which is achieved by a photon-quark
 interaction
\begin{equation}
\label{Lgammaqq}
\mathcal{L}^{(1)}_{{\gamma}qq} =
	- e~\mathcal{Q} \bar{\psi} \gamma^{\mu}\psi \mathcal{A}_{\mu}
\end{equation}
should be supplemented by the coupling of the photon field to pions and is
 given by the $\gamma\pi$ interaction Lagrangian
\begin{equation}
\label{Lgammapi}
\mathcal{L}^{(2)}_{\gamma\pi\pi} =
        -e~[{\pivec} \times \partial^{\mu}{\pivec} ]_3\mathcal{A}_{\mu}\ \ ,
\end{equation}
with additional contact interactions obtained by a gauge transformation of the
 pion field of Eq.~(\ref{Lpiqq}) we get
\begin{eqnarray}
\label{Tadpole1}
\mathcal{L}^{(1)}_{{\gamma\pi}qq}&=&
-e~\frac{f_{\pi qq}}{m_\pi}
\bar{\psi}\gamma^\mu\gamma_5 [\tauvec\times\pivec ]_3\psi\mathcal{A}_\mu \ \ ,\\
\label{Tadpole2}
\mathcal{L}^{(1)}_{{\gamma\pi\pi}qq} &=& 
-e~\frac{1}{4f_\pi^2}
\bar{\psi}\gamma^\mu [[\tauvec \times \pivec ] \times \pivec ]_3 \psi
 	 \mathcal{A}_\mu \ \ .
\end{eqnarray}
Considering Eqs.~(\ref{Lgammaqq} - \ref{Tadpole2}) at the one-pion loop level,
 the full set of diagrams responsible for the neutron charge form factor are
 shown in Fig.~\ref{fig:Currents}.  Since the bare quark core,
 Eq.~(\ref{Lgammaqq}), and the ``seagull'' terms ($\sim{\cal O}(1/M_n)$) shown
 in Fig.~\ref{fig:Currents}b and~\ref{fig:Currents}c do not contribute and
 since the tadpole contact graphs, Fig.~\ref{fig:Currents}e
 and~\ref{fig:Currents}f, cancel exactly,
the leading order contribution arises from two one-pion loop processes with
 $\gamma{qq}$ (Fig.~\ref{fig:Currents}d) and $\gamma\pi\pi$ couplings 
 (Fig.~\ref{fig:Currents}a), respectively.  
Following to the standard Feynmann
 rules the e.m. current operators responsible for Fig.~\ref{fig:Currents}a
 and~\ref{fig:Currents}d can be constructed and in the Breit frame, where
 $Q^2=\qvec^2$, the resulting Sachs electric form factors of the neutron can
 be obtained as
\begin{eqnarray}
\label{Core}
G^{\gamma qq}_{E_n} (\qvec^2) &=& \frac{\tilde{G}(\qvec^2)}{Z_2}
\sum_\alpha  \mathcal{K}_{N\alpha}   
\int 
	\frac{d^3\kvec}{(2\pi)^3}~\kvec^2 F^2(\kvec^2)
				~\mathcal{C}^{\gamma qq}_\alpha(\qvec,\kvec)
	\nonumber \\ \\
\label{Pion_current11}
G^{\gamma \pi \pi}_{E_n}(\qvec^2)&=& - \frac{F_\pi(\qvec^2)}{Z_2} \sum_\alpha
			\mathcal{K}_{N\alpha} \hspace*{10mm} \nonumber
			\\ 
&\times& \int \frac{d^3\kvec}{(2\pi)^3} F(\kvec^2) F({\kvec'}^2)
\mathcal{C}^{\gamma \pi \pi}_\alpha(\qvec,\kvec,\kvec') ~\kvec\cdot\kvec' 
\end{eqnarray}
where
$\mathcal{K}_{N\alpha}
=\left(f_{\pi{qq}}/m_\pi\right)^2C_{N\alpha}^2\chi^{\alpha}$
 and $\kvec'=\kvec+\qvec$.
The coefficients are calculated algebraically within the CQM and result in $\chi^N=2$, $\chi^\Delta=-4/9$, $C_{NN}=5/3$ and
 $C_{N\Delta}=2\sqrt{2}$.  
 $F_{\pi}(\qvec)$  is the pion
 e.m. form factor and $F({\kvec})$ is the nucleon axial form factor.
 Furthermore, $\tilde{G}(\qvec^2)$ in Eq.~(\ref{Core}) represents the electric form factor of the
 quark core.  At the one-pion loop level the e.m. current
 operators as provided by Fig.~\ref{fig:Currents} involve a four dimensional
 integration over the momentum $k$ of the virtual pions. The momentum
 dependent factors $\mathcal{C}^{\gamma\pi\pi}_\alpha(\qvec,\kvec,\kvec')$ and
 $\mathcal{C}^{\gamma qq}_\alpha(\qvec,\kvec)$ are the result of the
 integration over $k_0$, which we perform analytically by closing the contour
 in the lower half of the complex plane.
Eqs.~(\ref{Core}) and~(\ref{Pion_current11}) are exact and hold for any
 chiral quark model involving a cloud of pions. Instead of doing model
 calculations at this stage we proceed further by simplifying them.  It can be
 shown that in the limit of low $Q^2$, Eqs.~(\ref{Core})
 and~(\ref{Pion_current11}) can be reduced to
\begin{eqnarray}
\label{eq:gendc}
\left\{
\begin{array}{c}
G^{\gamma qq}_{E_n} (\qvec^2) \\ \\
G^{\gamma qq}_{E_n} (\qvec^2)
\end{array}
\right\}
=
\left[\sum_\alpha\xi_\alpha S_{\pi}^{\alpha} \right]
\left\{
\begin{array}{c}
\tilde{G}(\qvec^2) \\
- F_\pi(\qvec^2) F(\qvec^2)
\end{array}
\right\} 
\end{eqnarray}
where $S_{\pi}^{\alpha}$ is given by Eq.~(\ref{eq:genpi}) and the spin-isospin
 factors are $\xi_N=2/3$ and $\xi_\Delta=-1/3$.  Here the pion e.m. form
 factor~$F_{\pi}(\qvec)$, the nucleon axial form factor $F(\qvec)$ and the
 core electric form factor $\tilde{G}(\qvec^2)$ are the relevant ingredients
 which define the $Q^2$ behavior of the reduced form factors. The latter two
 are characteristics of the quark core. Their appearance is quite unique in
 all the quark models -- they do not exist in effective Lagrangian approaches
 at hadronic level.  The important feature of the CQM, at least for our
 considerations, is that the axial $F(\qvec^2)$ and electric
 $\tilde{G}(\qvec^2)$ form factors have the same functional form:
 $F(\qvec^2)$=$\tilde{G}(\qvec^2)$.  We use this fact to recombine
 Eqs.~(\ref{eq:gendc}) as
\begin{equation}
\label{en:total}
G_E^n(\qvec^2) =  \tilde{G}(\qvec^2) \left[ 1 - F_\pi(\qvec^2)\right] 
			\sum_\alpha\xi_\alpha{S^\alpha_\pi}
\end{equation}
Furthermore, as known from electroproduction experiments the pion
 e.m. form factor~$F_\pi(\qvec^2)$ can be fitted up
 to a few GeV~\cite{Bebek,Volmer} by a monopole term
\begin{equation}
\label{eq:pionff}
F_\pi(Q^2) = \Lambda_\pi^2/(\Lambda_\pi^2 + Q^2)
	   = (1+Q^2/\Lambda_\pi^2)^{-1}\ \ ,
\end{equation}
with a cut-off mass of $\Lambda_\pi^2$=0.53~GeV$^2$~\cite{Bebek,Volmer}.
Using Eq.~(\ref{eq:pionff}) 
the expression for $G_E^n$ can be written alternatively
\begin{eqnarray}
\label{eq:genfin}
G_E^n(Q^2) = \sum_{\alpha} \xi_{\alpha} S^{\alpha}_{\pi}\cdot
\left[\frac{Q^2/\Lambda_{\pi}^2}{1+Q^2/\Lambda_{\pi}^2}\right] \tilde{G}(Q^2)
 \ \ .
\end{eqnarray} 
We define the two parameters $a'$ and $b$ as 
\begin{equation}
\label{ba:deff}
b = {4M_n^2}/{\Lambda_{\pi}^2}~~~\mathrm{and}~~~
a'=\sum_\alpha\xi_{\alpha}S^\alpha_\pi, 
\end{equation}
If we assume in addition that the quark core electric
 form factor is given by the dipole ansatz, $\tilde{G}=G_D$, we then arrive at
 a Galster-like formulation for $G_E^n$ (cf. Eq.~\ref{eq:Galster})
\begin{equation}
\label{eq:ourgen}
G_E^n(Q^2) =  a'\frac{b\tau}{1 + b\tau}\ G_D(Q^2) \ \ .
\end{equation}
where $\tau=Q^2/4M_n^2$.  Already at this level ``our'' Galster form decouples
 the parameters responsible for shape (${b}$) and magnitude~(${a}'$), which
 reduces the uncertainty for $b$ by about a factor~2 in the fits. Note that
 the form of $\tilde{G}$ is model dependent; however, its generalization to
 other hadronic models is straightforward.  Presently we have used the dipole
 form to make the direct correspondence with the Galster ansatz.

$a'$ in Eqs.~(\ref{ba:deff}) and~(\ref{eq:ourgen}), which characterizes the
 spectroscopic strength of the pions in the nucleon, can be also fixed by
 imposing additional constraints coming from the experimental value of the
 neutron charge radius~\nrmsq. The latter is defined as the slope of
 $G_E^n(Q^2)$ for $Q^2\to0$, and applying this prescription to
 Eq.~(\ref{eq:genfin}) we obtain for
\begin{equation}
\nrmsq=-(6/\Lambda^2_\pi)\sum\xi_{\alpha}S^\alpha_\pi~~~\mathrm{or}~~~
{a}'=-  \nrmsq(\Lambda^2_\pi/6).
\end{equation}
Assuming \nrmsq\ and $F_\pi$ to be known, an even
 simpler expression for $G_E^n$ emerges
\begin{equation}
\label{eq:gensimple}
G_E^n(Q^2) = -\frac{\nrmsq}{6}\ Q^2 F_\pi(Q^2)\ G_D(Q^2)
\end{equation}
with no additional parameter. We mention that in Eq.~(\ref{eq:gensimple}) all
 quantities responsible for the properties of the pion cloud are
 experimentally accessible. This fact can be used to obtain the $G_E^n$ values
 directly from $\nrmsq$ and $F_{\pi}$.

\begin{figure}[t]
\begin{center} 
\includegraphics[clip=true,width=0.95\columnwidth,angle=0.]{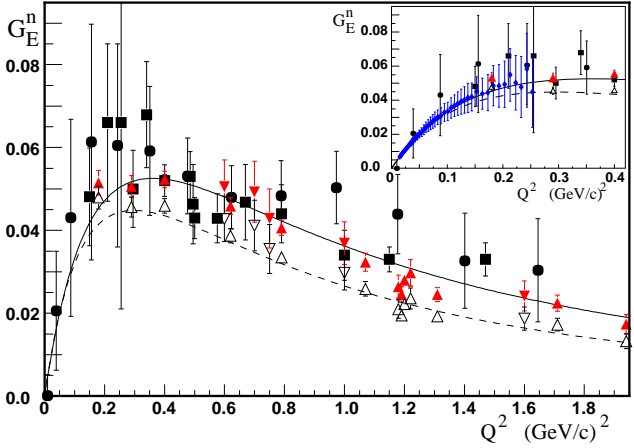}
\caption{\label{fig:gen}
The neutron electric $G_E^n$ form factor.
Squares: polarization data from Refs.~\cite{Becker}.
Circles:  elastic electron-deuteron data~\cite{Sick}.
Triangles: Eq.~\ref{eq:gensimple} with pion data  of Refs.~\cite{Bebek,Volmer};
 (open symbols: $\Lambda^2$=0.71~GeV$^2$, full: $\Lambda^2$=0.86~GeV$^2$).
 The insert shows the low $Q^2$ range with additional pion
 data~\cite{Amendolia}.
}
\end{center}
\end{figure}

Inserting the experimental values for \nrmsq~\cite{Kopecky} and for $F_\pi$ as
 extracted from electroproduction experiments~\cite{Bebek,Volmer} directly
 into the above equation we obtain the data (open triangles) plotted in
 Fig.~\ref{fig:gen}.  For comparison, $G_E^n$ values selected by
 Ref.~\cite{FriedrichW} are given by squares and the evaluation of
 Ref.~\cite{Sick} is presented by circles. The remarkable feature of the pion
 data is that they have smaller errors and that they lead to slightly smaller
 values for $G_E^n$ above $Q^2$=0.3~GeV$^2$.
In Fig.~\ref{fig:gen} the full curve represents the fit to the $G_E^n$
 data~\cite{FriedrichW,Becker} including the \nrmsq\ value, whereas the dashed
 curve represents the fit to the pion data with $\Lambda^2_D$=0.71~GeV$^2$
 with reasonable $\chi^2$ (Table~\ref{tab:fits}).
\begin{table}[b]
\begin{center}
\caption{\label{tab:fits}
Galster fits to $G_E^n$ data. $^\star)$ fixed in the fit.
}
\begin{tabular}{c|ccc|r}
\hline \hline
ref.&  $a'$ & $b$ & $\Lambda^2$/GeV$^2$ & $\chi^2$/ndof\\ \hline
\cite{FriedrichW,Becker}&0.369$\pm$0.008&4.69$\pm$0.11&0.71$^\star)$ &  9.9/14\\
\cite{Bebek}            &0.221$\pm$0.002&8.21$\pm$0.16&0.71$^\star)$ & 34.2/19\\
\cite{FriedrichW,Becker}&0.26$^\star)$&6.65$^\star)$ & 0.86$\pm$0.04 &  9.7/15\\
\hline
\end{tabular}
\end{center}
\end{table}

We note that, the use of $\Lambda_D^2 = 0.71$~GeV$^2$ in Eq.~(\ref{eq:ourgen})
 is not entirely correct as $\tilde{G}$ represents the core and not the
 extension of the total charge.  Using parameters $a'$ and $b$ according to
 the above formulae, a $\Lambda^2$ = 0.86~GeV$^2$ is obtained from a fit to
 the polarization data (third fit of Table~\ref{tab:fits}). The resulting
 curves from the first and third fit are identical below $Q^2$=2~GeV$^2$
 despite the different parameters.  Using this $\Lambda^2$=0.86~GeV$^2$, which
 corresponds to a $\sqrt{\langle{r^2}\rangle}$=0.74~fm of the core, results in
 derived $G_E^n$ values (full triangles in Fig.~\ref{fig:gen}) similar to the
 measured polarization data.  The insert show the obtained $G_E^n$ data for
 very low $Q^2$ range where more pion data~\cite{Amendolia} have been
 added. In this momentum transfer region $F_\pi$ can be understood to provide
 a model independent input to $G_E^n$. Since $\tilde{G}$ is close to unity,
 $G_E^n$ is considered to be parameter free at low $Q^2$.  Interestingly, the
 $G_E^n$ data from pion electroproduction fall right on top of the full curve.
 Clearly the pion data can only account for the soft contributions, however
 they give a lower limit for $G_E^n$ when using $\Lambda_D$.  Parameter $a'$
 shows that the weighted sum of pion spectroscopic factors is in the range of
 25\% (Table~\ref{tab:fits}).

In summary, in this work we have obtained a theoretical justification of the
 phenomenologically successful Galster parameterization of the neutron
 electric form factor $G_E^n$. We have shown that the chiral (pion cloud)
 content in the nucleon structure is the crucial ingredient that leads under
 some approximations to a Galster-like $Q^2$ dependence of $G_E^n$.  We found
 that all parameters of Eq.~(\ref{eq:ourgen}), i.e. $a'$ and $b$, can be fixed
 by other experiments and that they are connected with the existence of pions
 in the nucleons. The proposed parameterization of the $G_E^n$,
 Eqs.~(\ref{eq:ourgen}) and (\ref{eq:gensimple}), decouples the parameters
 responsible for shape (${b}$) and magnitude~(${a}'$) and allows to derive
 precise $G_E^n$ data at low $Q^2$ where the direct measurements of $G_E^n$ by
 double polarization experiments require larger corrections factors due to the
 notorious final state interactions.

We finally mention, that an analog prescription, Eq.~(\ref{eq:gensimple}), can
 be set up for the strange electric form factors and the strange radius of the
 nucleon.

\acknowledgement
{This work was supported by the Deutsche Forschungsgemeinschaft under contracts
  Gr1084/3 and GRK683, and LFSP(BW).}

\end{document}